# New Classification Methods for Hiding Information into Two Parts: Multimedia Files and Non Multimedia Files

Hamdan.O.Alanazi, A.A.Zaidan, B.B.Zaidan, Hamid A.Jalab and Zaidoon Kh. AL-Ani

*Abstract*— With the rapid development of various multimedia technologies, more and more multimedia data are generated and transmitted in the medical, commercial, and military fields, which may include some sensitive information which should not be accessed by or can only be partially exposed to the general users. Therefore, security and privacy has become an important, Another problem with digital document and video is that undetectable modifications can be made with very simple and widely available equipment, which put the digital material for evidential purposes under question .With the large flood of information and the development of the digital format Information hiding considers one of the techniques which used to protect the important information. The main goals for this paper, provides a general overview of the New Classification Methods for Hiding Information into Two Parts: Multimedia Files and Non Multimedia Files.

—————————— ◆ ——————————

## 1. INTRODUCTION

Due to advances in ICT, most of information is kept electronically. Consequently, the security of information has become a fundamental issue. Besides cryptography, steganography can be employed to secure information. Steganography is a technique of hiding information in digital media. In contrast to cryptography, the message or encrypted message is embedded in a digital host before passing it through the network, thus the existence of the message is unknown. Besides hiding data for confidentiality, this approach of information hiding can be extended to copyright protection for digital media: audio, video, and images [1].

- *Hamdan.O.Alanazi – Master Student, Department of Computer System & Technology, University Malaya, Kuala Lumpur, Malaysia.*
- *A. A. Zaidan – PhD Candidate on the Department of Electrical & Computer Engineering , Faculty of Engineering , Multimedia University , Cyberjaya, Malaysia.*
- *B. B. Zaidan – PhD Candidate on the Department of Electrical & Computer Engineering / Faculty of Engineering, Multimedia University, Cyberjaya, Malaysia.*
- *Dr. Hamid.A.Jalab- Senior Lecturer, Department of Computer Science &Information Technology, University Malaya, Kuala Lumpur, Malaysia*
- *Zaidoon Kh. AL-Ani - Master Student on the faculty of Information and Communication Technology at International Islamic University Malaysia, Kuala Lumpur, Malaysia.*

In addition, the rapid growth of publishing and broadcasting technology also require an alternative solution in hiding information. The copyright such as audio, video and other source available in digital form may lead to large-scale unauthorized copying. This is because the digital formats make possible to provide high image quality even under multi-copying. Therefore, The growing possibilities of modern communications need the special means of security especially on computer network. The network security is becoming more important as the number of data being exchanged on the Internet increases. Therefore, the confidentiality and data integrity are requires to protect against unauthorized access and use. This has resulted in an explosive growth of the field of information hiding [2].
the special part of invisible information is fixed in every image that could not be easily extracted without specialized technique saving image quality simultaneously .All this is of great concern to the music, film, book and software publishing industries[5],[7].
Information hiding is an emerging research area, which encompasses applications such as copyright protection for digital media, watermarking, fingerprinting, and steganography. All these applications of information hiding are quite diverse [8].

In watermarking applications, the message contains information such as owner identification and a digital time stamp, which usually applied for copyright protection [15].

Fingerprint, the owner of the data set embeds a serial number that uniquely identifies the user of the data set. This adds to copyright information to makes it possible to trace any unauthorized used of the data set back to the user [1].

Steganography hide the secret message within the host data set and presence imperceptible. In those applications, information is hidden within a host data set and is to be reliably communicated to a receiver. The host data set is purposely corrupted, but in a covert way, designed to be invisible to an informal analysis [21]. However, this paper will only focus on information hiding using steganography approach with in new classification methods for hiding information in to two parts: Multimedia File and Non Multimedia File [15].

## 2. HIDDEN IN MULTIMEDIA FILE

One of earlier and popular methods of hidden information is hidden data in the text file, it takes many phases and many ways, each one has its own affectivity and efficiency, some of these methods used for hidden data in the text as we shown in this paper how the German spy use this art, other technology use the space between the word to embed data.

### 2.1 Hidden in Text

The most common methods of concealment and simplest is Switches (Binary Digit) known briefly as (bit), least significant known as (LSB), Where it is altered binary digit characters to the message characters to be hidden, after conversion of such characters to byte as well as the American standard code for information interchange (ASCII). We cannot use this method here because switching binary digit might lead to an increase or decrease the value of letter by (1); this leads to the advance of this letter with the neighbor letter, for example the letter (C) in English represent in binary (100 0011) for instant if we replace the Least Significant bit the binary value become (100 0010) which is represent (B) in English, that will make the carrier text become a meaningless, which denies the goal of hidden technique, Therefore resort to other means which exploit spaces between strings and words in the carrier text [6].

Methods of concealment in the text are weak and inefficient it is also not suitable for the application, the main disadvantages include [13],[16]:

- Need large text to hide small message
- Some methods require complex techniques for the application like method of hidden information in the words [18].
- The possibility of noting changes in the carrier letter compared with the original
- The probability of crash the system of concealment in case of presenting the letter using one service applications such as (word) as a result of it contains some automatic format, which leads to change the long spaces between words (for the English version)[19].

Hidden techniques are limited for the text with difficulty concealing other types of messages like equations, charts, images and sounds.

### 2.1.1 Hidden in Space

In this technique; parts of the secret letter tight in to the spaces of the carrier letter these technical include two ways of methods; in the first method we may embed one or two spaces after each set of phrases for the carrier letters, and the second method contains the same way of the first method except that the process is embedding after every word of the carrier letter words [9].

The first method more efficient for lack of change compared with the original text but it is not efficient in terms of the amount of data that possible to embed, the way in the second method could embed more quantitative data but more likely for observation and perception compared to the original text [10].

### 2.1.2 Hidden in Words

This technique was used during the Second World War, sending a hidden message contained by another message which is not relevant, the role of this idea is nomination letters in every word of the carrier letter from the characters confidential letter to be sent [9].
One of these examples was the letter that was sent by Deutsch Spy during the Second World War "Apparently neutrals protest is thoroughly discounted and ignored. Is man hard hit Blockade issue affects pretext for embargo on by-products, ejecting suet's and vegetable oils" by analyzing this letter crafts taking the second letter in each word it will show us the following secret letter "Perishing sails from NY June1 "[10].

This method is very complex and not efficient in terms of application in the computer system since it requires an expert system in the field of language through which to choose the suitable words and build the required sentence, as well as a huge database requires [11].

### 2.2 Hidden in Digital Images

In the computer images are representing as two dimensional array, this arrays include the pixels, in fact the number of bits in pixels represent the type of image; each one representing intensity lighting of pixel (the smallest unite in the image ) These values used by Raster to view on the screen the place posed by each of these values. The process of access each unit values of the image before scanning depends on the representation technical, depends on the number of byte which used to represent each pixel, and that what they call it the representation of image, now the representation of the image is the number of byte in the single pixel [6].

The most common representation is the 24-bit, on which 3-byte represent the color of the image in the single pixel, there is other representation the call it 8-bit, in which only one byte represent the color in the image, and each method representation has its own characteristics and determinants in the process of concealment, in this research we will focus on the image with the extension (JPG, BMP)[13].

### 2.2.1 Hidden in (24) Bit Image

In this type of images the representation of each pixel is 24-bit where each 3 byte represent basic colors Red, Green, and Blue (the Color intensity lighting) And the integration of the three illumination intensity is formed the required color (RGB) note that these percentages represent the true colors, and therefore they will be adopted directly by the scanner to show the color required for the corresponding location on the screen, which is why so of this kind of representation called (True Color) [16].

The use of this type of representation means that it is possible to represent 256 different level of color for each of the three colors, this means that changing the least significant bit for each color cannot be realized by the human eye because of the change will increase or decrease the intensity of the lighting color by (1) only from total (256), also we can replace the 2nd least significant bit without the possibility of recognition that change [16].

We can take this advantage of this feature to hide data within these areas and that by replacement bit of image by bit of data which should be hidden [16].

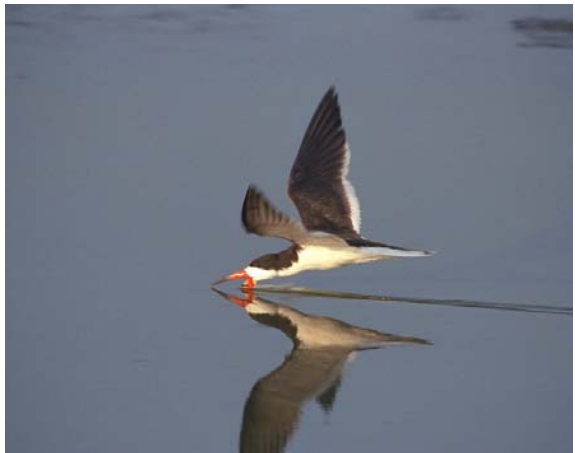

Fig 1. 24-bit images

### 2.2.2 Hidden in (8) Bit Image

In this type of images the presentation of each pixel by one byte or 8-bit, image is represent as a values in a matrix models which do not represent the true values of the image pixel, but it represent color index for the address on the schema just grandmothers of the true colors, or called (Palette) found in the image Header [18].

The number of colors level at different possible presentation in eight bit is (256) color level, namely the schema just grandmothers contains 256 colors different expect level of the color, there are two types of offers in this type of image one is represented by (256) different level of color (256 Color), and the other represented by (256) different level of intensity different gray scale, each of these types has its own characteristics and advantages [18].

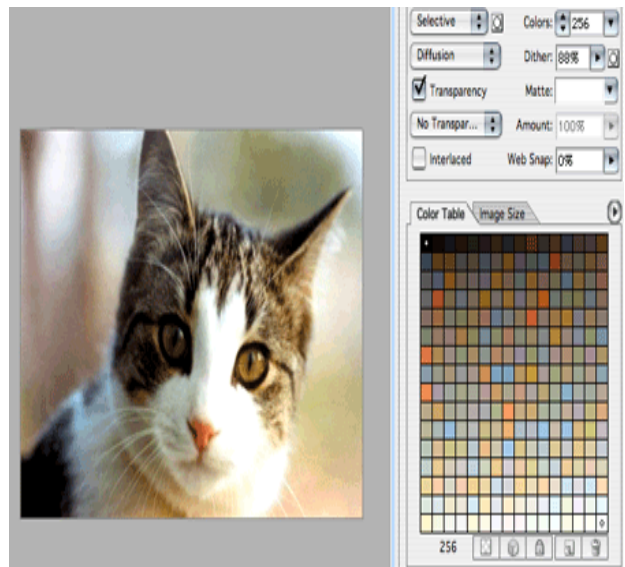

Fig 2. 8-bit images

### 2.2.3 Image Represent in 256 Colors

Here, 256 different 24 bit colors (can also use 16 or 32 bit) are selected out of the 16 million possible. These 256 colors are called the palette. Each pixel in the image is represented by a single byte. This byte is not the color but rather is the index of the color in the palette. When an image is stored, the 8 bit color index for each pixel is saved along with the palette. For a large image this can be significant space savings [18].

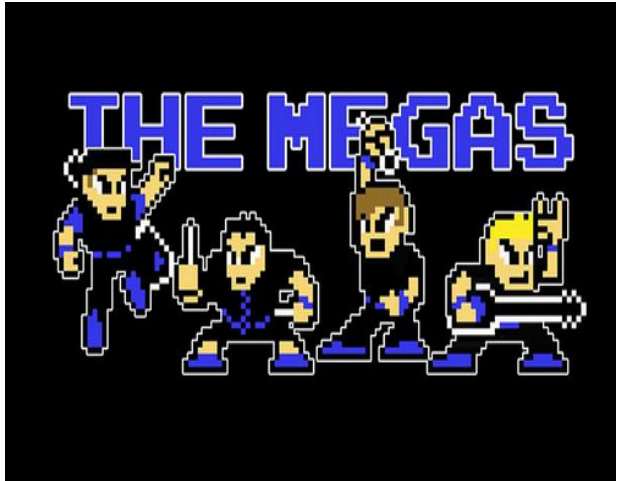

Fig 3. 8-bit color images

### 2.2.4 Image Represent by 256 of Grey Level

A grayscale or gray level image is simply one in which the only colors are shades of gray. The reason for differentiating such images from any other sort of color image is that less information needs to be provided for each pixel. In fact a `gray' color is one in which the red, green and blue components all have equal intensity in RGB space, and so it is only necessary to specify a single intensity value for each pixel, as opposed to the three intensities needed to specify each pixel in a full color image [19].

Often, the grayscale intensity is stored as an 8-bit integer giving 256 possible different shades of gray from black to white. If the levels are evenly spaced then the difference between successive gray levels is significantly better than the gray level resolving power of the human eye [19].

Grayscale images are very common, in part because much of today's display and image capture for hardware can only support 8-bit images. In addition, grayscale images are entirely sufficient for many tasks and so there is no need to use more complicated and harder-to-process color images [19].

### 2.3 Hidden Data in Wave Module

### 2.3.1 Hidden Data in the 8-bit Model Representation

It was pointed out that this pattern is represented by each audio sample size (1Byte/Sample) and common here that the process of concealment in last significant bit each model where the switch imperceptibly by authorizing human [4].

In the 8-bit audio file representation means that each model will be represented eight size, meaning that there is (256) audio level can be represented at the highest in this type is between (0 - 255) [4].

In this kind of representation researcher hide data in the first bit that less important to detect this concealment, that is to conceal ratio (12.5% from the size of the file)[4].

### 2.3.2 Hidden Data in the 16-bit Model Representation

Presentation of the audio sample here using two bytes for each sample (2 Byte / sample), a pattern that can be used as cover to hide in last two significant bit (2 LSB), or even in the third bit which is the less significant to achieve the same condition in the above. This is the kind of representation of each audio sample 16-bit, meaning that the number of audio levels that can be represented are (65536) a different level, so it is obvious that researcher find space to conceal much more than the previous type, for example, means changing, the bit less important, by increasing or decreasing one level of the above, and 2 bit change means by increasing or decreasing three levels of those worst off, and so on [4].

### 2.3.3 Color System

- **RGB**

The RGB color model is an additive color model in which red, green, and blue light are added together in various ways to reproduce a broad array of colors. The name of the model comes from the initials of the three additive primary colors, red, green, and blue [5].

The main purpose of the RGB color model is for the sensing, representation, and display of images in electronic systems, such as televisions and computers, though it has also been used in conventional photography. Before the electronic age, the RGB color model already had a solid theory behind it, based in human perception of colors [5].

- **Color**

Color or Colour is the visual perceptual property corresponding in humans to the categories called red, yellow, blue, black, etc. The RGB color model is the most common way to encode color in computing, and several different binary digital representations are in use. Color derives from the spectrum of light (distribution of light energy versus wavelength) interacting in the eye with the spectral sensitivities of the light receptors. Color categories and physical specifications of color are also associated with objects, materials, light sources, etc., based on their physical properties such as light absorption, reflection, or emission spectra [12].

- **Physical color**

Electromagnetic radiation is characterized by its wavelength (or frequency) and its intensity. When the wavelength is within the visible spectrum (the range of wavelengths humans can perceive, approximately from 380 nm to 740 nm), it is known as "visible light".Most light sources emit light at many different wavelengths; a source's spectrum is a distribution giving its intensity at each wavelength. Although the spectrum of light arriving at the eye from a given direction determines the color sensation in that direction, there are many more possible spectral combinations than color sensations. In fact, one may formally define a color as a class of spectra that give rise to the same color sensation, although such classes would vary widely among different species, and to a lesser extent among individuals within the same species. In each such class the members are called metamers of the color in question [3].

- **Color in bit**

The main characteristic of all of them is the quantization of the possible values per component (technically a sample) by using only integer numbers within some range [17].

## 3. HIDDEN IN NON MULTIMEDIA FILE

### 3.1 Hidden in Disk Space

Another way to hide information relies on hidden unused space that is not readily apparent to an observer. Taking advantage of an unused or reserved space to hold covert information provides a mean of hidden information without perceptually degrading the carrier. The way operating system stores files typically results in an unused space that appears to be allocated to files [15].

Another method of hidden information in file system is to create hidden partitions. These partitions are not seen if the system is started normally [15].

### 3.2 Hidden in Network Packets

With the rapid development of Internet technologies, the number of data packets sent and received electronically is increasing greatly. As the technology of transmitting information on network in secure, the importance of information hidden as a field of information security comes to be recognized widely [14].

Any of these packets can provide a covert communication channel. For example, TCP/IP packets are used to transport information over the internet. The pocket headers have unused space or other features that can be manipulated to hide information [21].

### 3.3 Hidden in Software and Circuitry

Data can also be hidden based on the physical arrangement of carrier. The arrangement itself may be an embedded signature that is unique to the creator. An example is in the layout of code distribution in a program or the layout of electronic circuits on a board. This type of 'marking' can be used to uniquely identify the design origin and can not be removed without significant change to the work [21].

Table 1
Weaknesses of Method for Hiding Information.

| Methods for Hiding Information. | Weakness |
|---|---|
| Hiding in Text | Methods of concealment in the text are weak and inefficient not suitable for the application and the main disadvantage include, need large text to hide small message that leads to an increase in the size of the cover. |
| Hiding in Disk Space | Methods of concealment in the unused areas weak and inefficient not suitable for the application and the main disadvantage include, this is not efficient since its easy detection by using some service software such as Norton antivirus. |
| Hiding in Network Packets | Methods of concealment in the network packets are weak and inefficient not suitable for the application and the main disadvantage include, Limitation of size for hiding information and It requires other methods to hide the data. |
| Hiding in Software and Circuitry | Methods of concealment in the (software or circuitry) are weak and inefficient not suitable for the application and the main disadvantage include, It is not possible to combine the (maximize/ maximum) strength of Robustness with maximize/maximum amount of hidden data comparing to data cover, which the hiding method cannot makes the relation between the cover and the message independent. |
| Hiding in Image and Audio | Methods of concealment in the (image or audio) are weak and inefficient not suitable for the application and the main disadvantage include:<br>a) The size of the output of the hidden data file is larger comparing to the encoded data. In its most efficient possible case, it may reach double the size of encoded data or a bet less .<br>b) In some situations output file may reach eight times larger than the encoded data, as well as certain files of media images, files may reach fifty times larger when they are encoded.<br>c) In the case of using a cover environment with equal value spaces as in the pictures with constant value colour spaces (week texture) or sounds with constant intensity sound intervals, that may lead to discovery or differentiation at these sectors. |

From the table above, it is clear that the main weaknesses of those methods are the size of the hidden data depends on the size of the cover file and any changes made are easily detected by anti virus software. Thus, through this paper, these weaknesses will be solved through the use of exe.file.

### 3.4 Hidden in Portable Executable File (PE-File)

The planned system uses a portable executable file as a cover to embed an executable program as an example for the planned system.This section is divided into four parts [1]:

- Characteristics of executable files
- Advantage of PE-file.

### 3.4.1 Characteristics of Executable Files

The characteristics of the Executable file does not have a standard size, like other files, for example the image file (BMP) the size of this file is between (2-10 MB), Other example is the text file (TXT) the size often is less than 2 MB.Through our study the characteristics of files have been used as a cover, it found that lacks sufficient size to serve as a cover for information to be hidden [15].

For these features of the Executable file, it has unspecified size; it can be 650 MB like window setup File or 12 MB such as installation file of multi-media players. Taking advantage of this feature (disparity size) make it a suitable environment for concealing information without detect the file from attacker and discover hidden information in this file [2].

### 3.4.2 Advantage of PE-File

The addition of the Microsoft® windows NT™ operating system to the family of windows™ operating systems brought many changes to the development environment and more than a few changes to applications themselves. One of the more significant changes is the introduction of the Portable Executable (PE) file format. The name "Portable Executable" refers to the fact that the format is not architecture specific .In other words, the term "Portable Executable" was chosen because the intent was to have a common file format for all flavors of windows, on all supported CPUs. The PE files formats drawn primarily from the Common Object File Format (COFF) specification that is common to UNIX® operating systems [5]. Yet, to remain compatible with previous versions of the MS-DOS® and windows operating systems, the PE file format also retains the old familiar MZ header from MS-DOS. The PE file format for Windows NT introduced a completely new structure to developers familiar with the windows and MS-DOS environments [7],[8]. Yet developers familiar with the UNIX environment will find that the PE file format is similar to, if not based on, the COFF specification .The entire format consists of an MS-DOS MZ header, followed by a real-mode stub program, the PE file signature, the PE file header, the PE optional header, all of the section headers, and finally, all of the section bodies[9],[10],[11],[15].

## 4. OTHER TECHNIQUES

Other techniques commonly used to hide data, using the unused areas of disk, for example in the operating system under Windows environment, organizes disk players (Formatted) using groupings (Clusters) size (32kB), and if the size of the file is open (1kB), the only remaining volume 31kB will be missed, and these could be used to conceal a significant amount of data[5],[12].

This is a good way in terms of size can take advantage of it to hide but it is not efficient since its easy detection by using some service software such as (Norton Commander) it could offer six-ten format, as well as the corresponding additional text note are hidden.

Other techniques and the use of abandoned areas (Reserved) of the prefix files (Files Header) image and sound files, and these methods also vulnerable to the possibility of distinguished, and the relatively small size [18],[19].

## 5. STRENGTH AND WEAKNESS FACTORS OF STEGANOGRAPHY

Although Steganography encompasses methods of transmitting secret messages through innocuous cover carriers in such a manner that the very existence of the embedded messages is undetectable, but it also has some weaknesses that attackers can utilize these weaknesses and detect and /or destruct the embedded message

### 5.1 Weaknesses in the Concealment Systems

The most important weaknesses of Steganography are as follows:
- The process of hidden data by using Least Significant Bit (LSB), which is a common method, used the image as a cover for data to be hidden, but it's mostly subjected to attacks from attackers. Once attackers suspect there is Steganography implemented in the data, they will reproduce LBS and examine it to check whether it has a meaning or not.
- the results of Steganography might be compressed by Lossy Compression to reduce the size of the file especially at transferring the data, which leads to destroy the hidden data in that file after the reopening
- In the case of using a cover environment with equal value spaces as in the pictures with constant value colour spaces (week texture) or sounds with constant intensity sound intervals, that may lead to discovery or differentiation at these sectors.
- One of the methods of breaking concealment in image and sound is changing the format of the file.

### 5.2 Disadvantages of the Concealment System

The disadvantages of Steganography are as follows:

- It requires other methods to hide the data.
- The size of the output of the hidden data file is larger comparing to the encoded data. In its most efficient possible case, it may reach double the size of encoded data or a bet less. In some situations output file may reach eight times larger than the encoded data, as well as certain files of media images and text, files may reach fifty times larger when they are encoded.

- It is not possible to combine the (maximize/ maximum) strength of Robustness with maximize/maximum amount of hidden data comparing to data cover.

## 6. CONCLUSION

In this paper a overview for the new classification methods for Hiding Information were presented into Two Parts: Multimedia Files , one of earlier and popular methods of hidden information are hidden data in the (text file or Digital Images or Wave Module ), it takes many phases and many ways, each one has its own affectivity and efficiency, Non Multimedia Files methods of hidden information are hidden data in the (Disk Space Network Packets, Software and Circuitry), in this paper explained, why the researcher used P-E-File as the cover file and it classifier to the non multimedia file , finally it discussed about Characteristics of Executable Files , Advantage of PE-File and Strength and Weakness Factors of Steganography.

## ACKNOWLEDGEMENT

This research was fully supported by "King Saud University", Riyadh, Saudi Arabia. The author would like to acknowledge all workers involved in this project whom had given their support in many ways, aslo he would like to thank in advance Dr. Musaed AL-Jrrah, Dr. Abdullah Alsbail, Dr. Abdullah Alsbait. Dr.Khalid Alhazmi, Dr.Abdullah Al-Afnan, Dr.Ibrahim Al-Dubaian and all the staff in king Saud University especially in Applied Medical Science In "Al-Majmah" for thier unlimited support, without thier notes and suggestion this research would not be appear.

## REFERENCES

[1] A.A.Zaidan, B.B.Zaidan, Fazidah Othman, "New Technique of Hidden Data in PE-File with in Unused Area One", International Journal of Computer and Electrical Engineering (IJCEE), Vol.1, No.5, ISSN: 1793-8198, p.p 669-678.

[2] A.A.Zaidan, B.B.Zaidan, Anas Majeed, "High Securing Cover-File of Hidden Data Using Statistical Technique and AES Encryption Algorithm", World Academy of Science Engineering and Technology (WASET), Vol.54, ISSN: 2070-3724, P.P 468-479.

[3] A.A.Zaidan, B.B.Zaidan, "Novel Approach for High Secure Data Hidden in MPEG Video Using Public Key Infrastructure", International Journal of Computer and Network Security, 2009, Vol.1, No.1, ISSN: 1985-1553, P.P 71-76.

[4] Mohamed A. Ahmed, Miss Laiha Mat Kiah, B.B.Zaidan, A.A.Zaidan,"A Novel Embedding Method to Increase Capacity of LSB Audio Steganography Using Noise Gate Software Logic Algorithm", Journal of Applied Sciences, Vol.10, Issue 1, ISSN: 1812-5654, 2010, P.P 59-64

[5] A.W.Naji, A.A.Zaidan, B.B.Zaidan, Shihab A, Othman O. Khalifa, " Novel Approach of Hidden Data in the (Unused Area 2 within EXE File) Using Computation Between Cryptography and Steganography ", International Journal of Computer Science and Network Security (IJCSNS) , Vol.9, No.5 , ISSN : 1738-7906, pp. 294-300.

[6] Anas Majed Hamid, Miss Laiha Mat Kiah, Hayan .T. Madhloom, B.B Zaidan, A.A Zaidan," Novel Approach for High Secure and High Rate Data Hidden in the Image Using Image Texture Analysis", International Journal of Engineering and Technology (IJET) , Published by: Engg Journals Publications, ISSN:0975-4042, Vol.1,NO.2,P.P 63-69.

[7] A.A.Zaidan, Fazidah. Othman, B.B.Zaidan, R.Z.Raji, Ahmed.K.Hasan, and A.W.Naji," Securing Cover-File without Limitation of Hidden Data Size Using Computation between Cryptography and Steganography ", World Congress on Engineering 2009 (WCE), The 2009 International Conference of Computer Science and Engineering, Proceedings of the International Multi Conference of Engineers and Computer Scientists 2009, ISBN: 978-988-17012-5-1, Vol.I, p.p259-265.

[8] A.A.Zaidan, A.W. Naji, Shihab A. Hameed, Fazidah Othman and B.B. Zaidan, " Approved Undetectable-Antivirus Steganography for Multimedia Information in PE-File ",International Conference on IACSIT Spring Conference (IACSIT-SC09) , Advanced Management Science (AMS), Listed in IEEE Xplore and be indexed by both EI (Compendex) and ISI Thomson (ISTP),  Session 9,  P.P 425-429.

[9] A.A.Zaidan, B.B.Zaidan, M.M.Abdulrazzaq, R.Z.Raji, and S.M.Mohammed," Implementation Stage for High Securing Cover-File of Hidden Data Using Computation Between Cryptography and Steganography", International Conference on Computer Engineering and Applications (ICCEA09), Telecom Technology and Applications (TTA), indexing by Nielsen, Thomson ISI (ISTP), IACSIT Database, British Library and EI Compendex, Vol.19, Session 6, p.p 482-489.

[10] A.W. Naji, A.A.Zaidan, B.B.Zaidan, Ibrahim A.S.Muhamadi, "New Approach of Hidden Data in the portable Executable File without Change the Size of Carrier File Using Distortion Techniques", Proceeding of World Academy of Science Engineering and Technology (WASET),Vol.56, ISSN:2070-3724, P.P 493-497.

[11] A.W. Naji, A.A.Zaidan, B.B.Zaidan, Ibrahim A.S.Muhamadi, "Novel Approach for Cover File of Hidden Data in the Unused Area Two within EXE File Using Distortion Techniques and Advance Encryption Standard.", Proceeding of  World Academy of Science Engineering and Technology (WASET),Vol.56, ISSN:2070-3724, P.P 498-502.

[12] M. Abomhara, Omar Zakaria, Othman O. Khalifa , A.A.Zaidan, B.B.Zaidan, "Enhancing Selective Encryption for H.264/AVC Using Advance Encryption Standard ", International Journal of Computer and Electrical Engineering (IJCEE), ISSN: 1793-8198,Vol.2 , NO.2, April 2010, Singapore..

[13] Md. Rafiqul Islam, A.W. Naji, A.A.Zaidan, B.B.Zaidan " New System for Secure Cover File of Hidden Data in the Image Page within Executable File Using Statistical Steganography Techniques", International Journal of Computer Science and Information Security (IJCSIS), ISSN: 1947-5500, P.P 273-279, Vol.7 , NO.1, January 2010, USA..

[14] Hamid.A.Jalab, A.A Zaidan, B.B Zaidan, "New Design for Information Hiding with in Steganography Using Distortion Techniques", International Journal of Engineering and Technology (IJET)), Vol 2, No. 1, ISSN: 1793-8236, Feb (2010), Singapore.

[15] Hamdan. Alanazi, Hamid.A.Jalab, A.A.Zaidan, B.B.Zaidan, "New Frame Work of Hidden Data with in Non Multimedia File", International Journal of Computer and Network Security, 2010, Vol.2, No.1, ISSN: 1985-1553, P.P 46-54,30 January, Vienna, Austria.

[16] A.A.Zaidan, B.B.Zaidan, Hamid.A.Jalab," A New System for Hiding Data within (Unused Area Two + Image Page) of Portable Executable File using Statistical Technique and Advance Encryption Standared ", International Journal of Computer Theory and Engineering (IJCTE), 2010, VOL 2, NO 2, ISSN:1793-8201, Singapore.

[17] Alaa Taqa, A.A Zaidan, B.B Zaidan ,"New Framework for High Secure Data Hidden in the MPEG Using AES Encryption Algorithm", International Journal of Computer and Electrical Engineering (IJCEE),Vol.1 ,No.5, ISSN: 1793-8163, p.p.566-571 , December  (2009). Singapore.

[18] B.B Zaidan , A.A Zaidan ,Alaa Taqa , Fazidah Othman ," Stego-Image Vs Stego-Analysis System", International Journal of

Computer and Electrical Engineering (IJCEE),Vol.1 ,No.5 , ISSN: 1793-8163, pp.572-578 , December (2009), Singapore.

[19] A.W. Naji, Shihab A. Hameed, B.B.Zaidan, Wajdi F. Al-Khateeb, Othman O. Khalifa, A.A.Zaidan and Teddy S. Gunawan, " Novel Framework for Hidden Data in the Image Page within Executable File Using Computation between Advance Encryption Standared and Distortion Techniques", International Journal of Computer Science and Information Security (IJCSIS), Vol. 3, No 1 ISSN: 1947-5500, P.P 73-78,3 Aug 2009, USA.

[20] Hamid.A.Jalab, A.A Zaidan and B.B Zaidan," Frame Selected Approach for Hiding Data within MPEG Video Using Bit Plane Complexity Segmentation", Journal of Computing (JOC), Vol.1, Issue 1, ISSN: 2151-9617, P.P 108-113, December 2009, Lille, France.

[21] Mahmoud Elnajjar, A.A Zaidan, B.B Zaidan, Mohamed Elhadi M.Sharif and Hamdan.O.Alanazi ," Optimization Digital Image Watermarking Technique for Patent Protection", Journal of Computing (JOC), Vol.2, Issue 2, ISSN: 2151-9617, P.P 142-148 February 2010, Lille, France.

**Hamdan Al-Anazi**: has obtained his bachelor dgree from "King Suad University", Riyadh, Saudi Arabia. He worked as a lecturer at Health College in the Ministry of Health in Saudi Arabia, then he worked as a lecturer at King Saud University in the computer department. Currently he is Master candidate at faculty of Computer Science & Information Technology at University of Malaya in Kuala Lumpur, Malaysia. His research interest on Information Security, cryptography, steganography and digital watermarking, He has contributed to many papers some of them still under reviewer.

**Aos Alaa Zaidan**: He obtained his 1st Class Bachelor degree in Computer Engineering from university of Technology / Baghdad followed by master in data communication and computer network from University of Malaya. He led or member for many funded research projects and He has published more than 55 papers at various international and national conferences and journals, His interest area are Information security (Steganography and Digital watermarking), Network Security (Encryption Methods) , Image Processing (Skin Detector), Pattern Recognition , Machine Learning (Neural Network, Fuzzy Logic and Bayesian) Methods and Text Mining and Video Mining. .Currently, he is PhD Candidate on the Department of Electrical & Computer Engineering / Faculty of Engineering / Multimedia University / Cyberjaya, Malaysia. He is members IAENG, CSTA, WASET, and IACSIT. He is reviewer in the (IJSIS, IJCSN, IJCSE and IJCIIS).

**Bilal Bahaa Zaidan:** He obtained his bachelor degree in Mathematics and Computer Application from Saddam University/Baghdad followed by master in data communication and computer network from University of Malaya. He led or member for many funded research projects and He has published more than 55 papers at various international and national conferences and journals, His interest area are Information security (Steganography and Digital watermarking), Network Security (Encryption Methods) , Image Processing (Skin Detector), Pattern Recognition , Machine Learning (Neural Network, Fuzzy Logic and Bayesian) Methods and Text Mining and Video Mining. .Currently, he is PhD Candidate on the Department of Electrical & Computer Engineering / Faculty of Engineering / Multimedia University / Cyberjaya, Malaysia. He is members IAENG, CSTA, WASET, and IACSIT. He is reviewer in the (IJSIS, IJCSN, IJCSE and IJCIIS).

**Dr.Hamid.A.Jalab**: Received his B.Sc degree from University of Technology, Baghdad, Iraq. MSc & Ph.D degrees from Odessa Polytechnic National State University 1987 and 1991, respectively. Presently, Visiting Senior Lecturer of Computer System and Technology, Faculty of Computer Science and Information Technology, University of Malaya, Malaysia. various international and national conferences and journals, His interest area are Information security (Steganography and Digital watermarking), Network Security (Encryption Methods) , Image Processing (Skin Detector), Pattern Recognition , Machine Learning (Neural Network, Fuzzy Logic and Bayesian) Methods and Text Mining and Video Mining.

**Zaidoon Kh. AL-Ani: H**as gained his bachelor Degree in computer science from University of Baghdad in 2003. Presently, he is conducting his master degree in Information and Communication Technology at International Islamic University Malaysia. The field of interest is steganography. Projects have been done which are titled as "Data security using hiding data" and Evaluation of Steganography for Arabic text".